\documentclass[prb, lengthcheck, groupedaddress, showpacs, showkeys, letterpaper, reprint]{revtex4-1}

\usepackage{amssymb,amsmath,xcolor,graphicx,xspace,colortbl,ragged2e,rotating} %
\usepackage{amsfonts}  
\usepackage{amsmath}  
\usepackage{amssymb}  
\usepackage{boxedminipage}  
\usepackage{graphics}  
\usepackage{graphicx}  
\usepackage{ragged2e}  
\usepackage{wrapfig}  
\usepackage{xcolor}  
\graphicspath{{PRBRCFinalAccepted121218_graphics/}{PRBRCFinalAccepted121218_tcache/}{PRBRCFinalAccepted121218_gcache/}}
\DeclareGraphicsExtensions{.pdf,.eps,.ps,.png,.jpg,.jpeg}
\providecommand{\U}[1]{\protect\rule{.1in}{.1in}}

\begin{document}
\preprint{HEP/123-qed
}
\title[Acoustic Parametric Oscillator
]{Evidence for terahertz acoustic phonon parametric oscillator based on
acousto-optic degenerate four-wave mixing in silicon doping superlattice
}
\author{Thomas E. Wilson
}
\affiliation{Department of Physics, Marshall University, Huntington, WV 25755
}
\begin{abstract}We report first evidence for a 1.0 terahertz (THz) self-starting mirrorless acoustic
phonon parametric oscillator (MAPPO) produced from acousto-optic phase-conjugate
degenerate four-wave (D4WM) mixing in a THz laser-pumped silicon doping superlattice (DSL). The DSL
was grown by molecular beam epitaxy on a (100) boron-doped silicon substrate. A superconducting
NbTiN subwavelength grating was used to couple the THz laser radiation into the DSL. Superconducting
granular aluminum bolometric detection, coupled with Si:B piezophonon spectroscopy,
revealed excitation of THz coherent compressional and
shear waves, along the \textless{}111\textgreater{} direction only. The Bragg scattering condition for distributed
feedback, and the energy conservation requirement for the D4WM\ process, were both verified. A THz MAPPO could provide a testbed for
studies of non-classical acoustic phonon fields.
\end{abstract}
\volumeyear{year
}
\volumenumber{number
}
\issuenumber{number
}
\eid{identifier
}
\date{date}
\received[Received text]{date
}
\revised[Revised text]{date
}
\accepted[Accepted text]{date
}
\published[Published text]{date
}
\startpage{101
}
\endpage{102
}
\maketitle
Phonons, the quanta associated with the vibrational waves of a crystalline lattice,
are among the most important excitations in condensed matter. Major advances in our understanding
of the physics of phonons during the past several decades have included first principles
calculations of the phonon lifetimes in simple crystals, their interactions
with other phonons, defects, boundaries and impurities, and in controlling thermal transport at the
nanoscale\ \cite{Cahill}. A
coherent source of acoustic phonons could find widespread applications in science and technology.
The stimulated emission of transverse acoustic (TA)  phonon avalanches at 50.4 gigahertz (GHz) has
been observed in dilute centers in ruby \cite{Tilstra}. Current phonon studies use visible-light pulse (VLP)
transducer techniques \cite{Ruello} to produce
and detect coherent near-THz acoustic phonons. In most VLP-based experiments, the generation
of coherent phonons was limited to longitudinal acoustic (LA) phonons because of the lateral
symmetry of the excited crystal, although TA\ phonons at 0.4 THz
have also been produced by piezoelectric coupling
in GaN multiple quantum wells \cite{Chen}. A (femto-
or picosecond) VLP produces, by electronic excitation and relaxation, a stress as a source
of near-THz coherent acoustic phonons. Stress joined with an LA or TA wave can be monitored
by a pump-probe technique measuring changes in reflectivity. A VLP transducer can consist of a metal
film, a surface region of a crystal, or a superlattice
(SL). More recently, the sound amplification by stimulated emission of radiation (saser) in
semiconductor superlattice structures has been described in connection with the VLP transducer
technique. One type of saser uses an active medium, a population inversion in a voltage-biased SL
containing free electrons \cite{Maryam,Kent2}.
Gain is due to phonon stimulated electronic transitions between next nearest energy levels of a
Wannier-Stark ladder; the gain coefficient for LA waves at 650 GHz was
$3 \times 10^{3}$
cm$^{ -1}$
. Another type \cite{Shinokita} utilizes a miniband
current in a SL (at room temperature) showing gain due to the acousto-electric effect; its gain
coefficient for LA waves near 400 GHz was
$8 \times 10^{3}$
cm$^{ -1}$
.

In this Rapid Communication, we report the first experimental evidence for a
mirrorless acoustic phonon parametric oscillator (MAPPO) in a doping superlattice structure
operating at one terahertz. It is based upon acousto-optic (AO) phase-conjugate degenerate four-wave
mixing (4WM)\cite{Yariv and Pepper} of counter-propagating fields.
The MAPPO is optically-pumped by nano-second (ns)\  pulsed 1.04 THz
plane-polarized laser radiation. Coherent light sources based upon 4WM in mirrorless laser
oscillators, in both hot and cold atoms \cite{Mei}, are known to have some unique properties such as cavity-free alignment and large
tunability. We have designed the SL period to provide spatially-distributed feedback. Distributed
feedback occurs when the Bragg scattering condition is satisfied for the signal and phase-conjugate
acoustic waves propagating normal to a gain grating \cite{Brignon}
associated with the 4WM process. We have not however, made any first-principles
calculations, and more data is necessary to further validate, and more fully understand the
phenomenon.

A ``n-i-p-i'' DSL  \cite{Dohler1} is formed by a periodic
$n\text{-}$
and
$p$
-doping structure in an otherwise homogeneous semiconductor crystal with undoped ($i\text{-}$
) regions in between. In its ground state the sample thus consists of a single crystal (in
contrast to compositional SLs) with alternately positively and negatively charged sheets of ions.
Given the backfolded acoustic phonons   (see Fig. 1 of Ruden and Dohler\cite{Ruden}) under the superlattice aspect are optical
modes, doping SLs
could be expected to be excitable by FIR\ electromagnetic fields. Our delta-doped doping
SLs were grown by MBE on (100) float-zone
$150$
mm diameter silicon wafers and consisted of\
$30$
sequential \textit{n-i-p-i'}s with period
$d =8.1$
nm        and dopant ($B$
and
$S b$) concentrations: N$\text{}_{A}^{2 D}$
= N$\text{}_{D}^{2 D}$
=
$3.3$
$ \times $
$10^{13}$
cm$^{ -2}$
. The smallest subband energy level spacing in the V-shaped potential wells for both
electrons and holes is of order 50 meV in our doping
SL \cite{Gossmann}, much
larger than the 4.3 meV pump radiation. Intense THz fields on the order of MV/cm have been reported
to induce carrier generation by impact ionizations
associated with impurity states in undoped multiple-quantum wells\cite{Tanaka}, however, in our experiments we use quite
modest THz fields on the order 10 V/cm. Accordingly our AO 4WM process should lie in the
reactive\ regime \cite{Flytzanis} where the FIR light
frequency is below the onset of absorption by electronic transitions.

Acoustic phase conjugation has been observed in piezoelectric semiconductors at
microwave frequencies\cite{Fossheim} resulting from a nonlinear piezoelectric coupling between
photons and phonons (or electric field
$E$
and elastic strain
$S$
) of the form
$\gamma E^{2}S^{2}$, with
$\gamma $
the coupling constant. This four-wave coupling is of the same form as that found in optical
4WM giving rise to optical phase conjugation. We postulate that electrostriction
coupling of this form is present in our experiments due to the direct action of the electric
field acting upon the ion sheets of the  DSL. Recently, electrostriction has also been reported to have produced an extraordinarily
strong D4WM response yielding THz acoustic breathing modes in gold nanoparticles
\cite{Xiang}.  Figure 1 illustrates
schematically our posited 1.04 THz MAPPA\  process (energy conservation
and distributed feedback have been verified as described below). In AO D4WM one of the optical pump
waves and the acoustic
probe wave couple to produce an index grating \cite{Rogovin}. This grating coherently scatters the
counter-propagating optical pump wave to produce a phase-conjugate acoustic wave. Distributed
feedback coupled with parametric gain, can then resonantly generate a
standing-wave pattern along \textless{}111\textgreater{}. The index gain-grating planes are normal to
the acoustic wavevector.  In contrast to OPOs, for MAPPOs the difficulties of phase
matching due to dispersion, are largely circumvented \cite{Flytzanis}.\
\begin{figure}\centering 
\setlength\fboxrule{0in}\setlength\fboxsep{0.1in}\fcolorbox[HTML]{FFFFFF}{FFFFFF}{\includegraphics[ width=2.7049180327868854in, height=2.5in,]{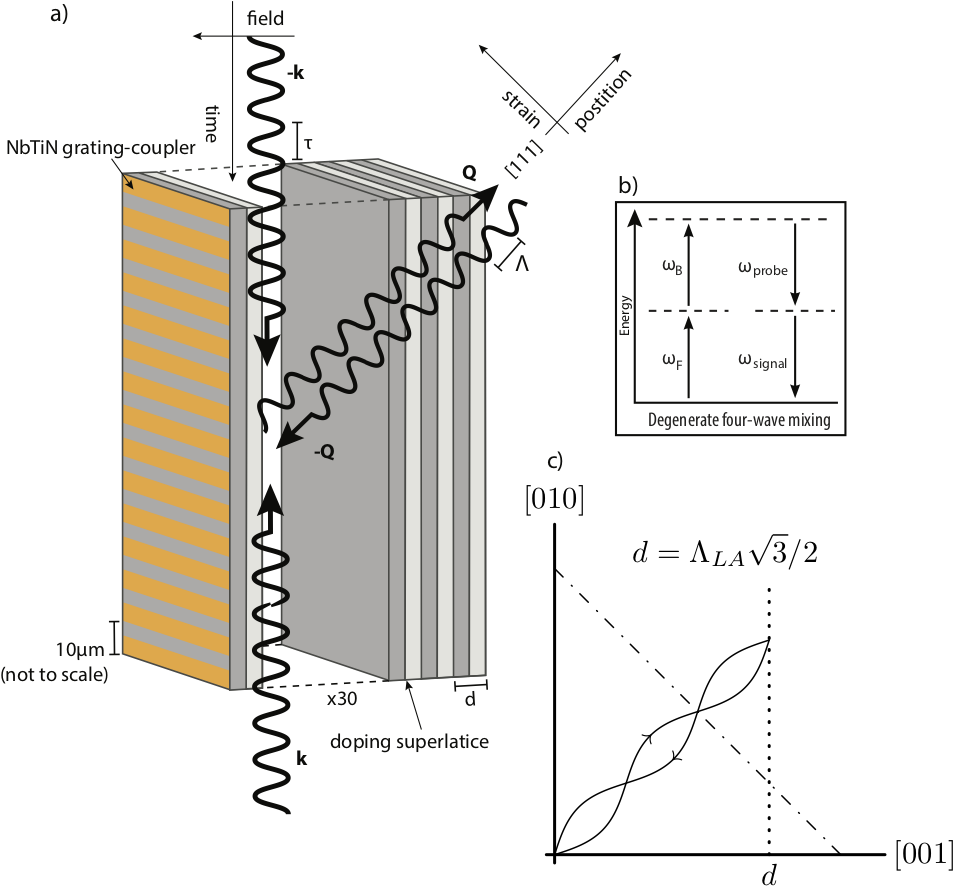}
}
\caption{(a) Schematic diagram of MAPPO showing k/-k evanescent
E\&M\ fields, and \textless{}111\textgreater{}-directed Q/-Q (PC) acoustic
fields. (b) Energy conservation diagram for D4WM (c)\ Distributed-feedback phase-matching, projected onto
$(100)\text{  plane, results in a}$
standing LA wave (solid lines) for Q/-Q.
$\text{}$
DSL
(dotted line) of period
$d$
. Gain grating (dash-dotted line) shown. LA wavelength:
$\Lambda _{LA}$}\label{Figure 1}\end{figure}

Figure 2 illustrates our experimental arrangement.
\begin{figure}\centering 
\setlength\fboxrule{0in}\setlength\fboxsep{0.1in}\fcolorbox[HTML]{FFFFFF}{FFFFFF}{\includegraphics[ width=2.541666666666667in, height=2.65625in,]{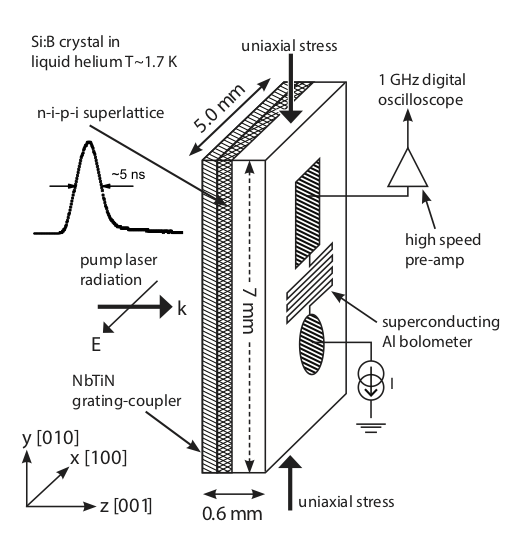}
}
\caption{Schematic diagram of experimental arrangement. Thicknesses of grating-coupler
and  DSL
(cross-hatched) not to scale. THz pump laser radiation incident from left with wavevector
k, E-field, and typical digitized pulse envelope shown.
}\end{figure}
A silicon crystal (size 7 mm
$ \times $
5 mm
$ \times $
0.60 mm) is immersed in vapor-pumped liquid helium ($1.66$
K) and carries on one surface a superconducting NbTiN grating-coupler
patterned on top of the doping SL, and on the opposite
crystal surface a superconducting Al bolometer. The grating-coupler is illuminated by
nanosecond-pulsed polarized
$1.04$
THz laser radiation. We chose to use NbTiN for the grating-coupler ($500\text{}$
nm thick, 10
$\mu $m period, 6
$\mu $m line, 4
$\mu $m space) due to its high frequency gap
$1.2$
THz \cite{Leone1} and low absorptivity, to preclude the
generation of broadband incoherent acoustic phonons. The subwavelength
grating-coupler converts the incident FIR electromagnetic wave into counter-propagating evanescent
surface waves within the DSL
as shown in Fig. 1(a). Grating-coupler theory \cite{Li and McCombe} shows that 85\% of the incident FIR laser intensity is converted into the surface
waves with 15\% remaining in the propagating field. The evanescent field, with a 1.7
$\mu $m decay length, can be assumed to be uniform over the
$0.24$
$\mu $m thick DSL, and does not extend appreciably into the
$0.60$
mm thick Si:B substrate.      The
$100$
nm thick, serpentine bilayer granular aluminum/palladium bolometer \cite{Wilson1} with active area
$A =$
$0.53$
mm$^{2}$
, was used for the detection of both photons and phonons. The phase-conjugate acoustic wave
would not be detected. The bolometer's responsivity
$\mathcal{R} =900$
V/W, was obtained through a combined dc, and ns-pulsed, characterization study
\cite{Danilchenko1}. The
bolometer was biased at a constant current of
$40$
$\mu $A, and the voltage signal was amplified
$40 \times $
using
$1$-GHz bandwidth electronics, and recorded using a fast real-time digital sampling
oscilloscope.

The Si substrate was boron-doped at a density of
$N_{B} =10^{15}$
cm$^{ -3}$
in order to use Si:B piezo-phonon spectroscopy \cite{Berberich and Schwarte 1, Berberich and Kinder}. The resonant phonon scattering by the stress-split
acceptor ground state can be used for phonon spectroscopy. We chose the boron doping density
$N_{B}$
appropriate for our sample thickness, according to an scaling law found in the Appendix of
Schwarte and Berberich \cite{Schwarte and Berberich}:
$N_{B} \sim N_{SB}\frac{4mm}{0.6mm}$
where
$N_{SB} =4.8 \times 10^{13}$
cm\textsuperscript {-3} and $4$
mm, were their doping level and sample thickness, respectively. The scattering rate for
TA\ phonons was shown\cite{Schwarte and Berberich} to be larger than for
LA\ phonons, although the respective scattering rates approach
each other near 5 meV. We used a uniaxial stress apparatus designed after
Fr{\"o}hlich and Nieswand \cite{Froelich1}, and
calibrated up to
$3$
kbar to apply the stress via a spring balance mounted above the cryostat.

We used a previously described \cite{Wilson2} custom cavity-dumped, optically-pumped, FIR molecular gas
laser to pump the DSL. The FIR laser resonator, filled with methyl fluoride at a pressure of
$6$
Torr and pumped by the 9R20 line of a multi-mode TEA\ CO\textsubscript {2}
laser, yielded smooth cavity-dumped pulse envelopes [$ \Delta t$
$ \simeq $
$5$
ns (FWHM)] (see Fig.2), as measured by a 1-GHz bandwidth pyroelectric detector, containing
the linearly-polarized
$1.04$
THz radiation. Peak powers of order
$10$
kW were obtained at
$10$
Hz pulse repetition rates. The measured wavelength,
$\lambda  =288.2 \pm 0.05$
$\mu $m, obtained with a scanning metal-mesh ($1000$
lpi) Fabry-Perot inteferometer \cite{Renk2}, agreed with the published wavelength \cite{Gross} for this FIR\ line. The energy contained in a
FIR\ pulse fluctuated by$ \approx 50 \%$
from pulse to pulse, with frequent larger excursions. A waveguide \cite{Woskov1}, used for FIR\ laser beam transport
to the entrance window of the liquid helium cryostat, was interrupted to insert a pair of crossed
wire-grid polarizers to reduce the pump power to
$ \sim 10$
mW.

Figure 3 shows a stacked sequence of digitized bolometer traces (signal versus time)
in response to single laser pulses of progressively increasing power. Due to the nonlinearity of the
bolometer response (described below) however, the listed powers are suspect. The first pulse
observed (time
$0$
ns) in all traces is the bolometer response to the arrival of undiffracted grating-coupled
FIR\ laser radiation. As the
laser radiation pump power reaches a threshold power of
$6.4$
mW, ballistic acoustic phonons appear at a delay time of
$110 \pm 2$
ns. As the incident FIR\ power increases further, the
acoustic signal at the
$110$
ns delay grows progressively stronger, and at a
$8.8$
mW pump power, a second phonon signal appears at
$210 \pm 2$
ns.       The distinct onsets of the phonon signals with pump power may be misleading however
due to the nonlinearity of the bolometer response. 
\begin{figure}\centering 
\setlength\fboxrule{0in}\setlength\fboxsep{0.1in}\fcolorbox[HTML]{FFFFFF}{FFFFFF}{\includegraphics[ width=2.988792in, height=1.9925279999999999in,]{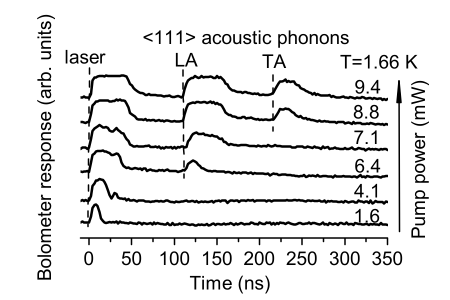}
}
\caption{Stacked sequence of digitized bolometer signals (arb. units)
for increasing (estimated) pump power (mW) listed above traces.
}\label{Figure 2}\end{figure}
For determination of group velocities of waves consistent with observed signals, we use the
equations of motion (Christoffel's equations \cite{Auld}) for
sound waves in silicon (mass density
$\rho  =$
$2.332$
g/cm$^{3}$
, elastic constants
$c_{1 1} =167.8$
GPa,
$c_{1 2} =65.2$
GPa,
$c_{4 4} =$
$80.01$
GPa) \cite{BerkeMayerWehner}. Among the surveyed modes for
\textless{}100\textgreater{}, \textless{}110\textgreater{}, \textless{}111\textgreater{}, \textless{}012\textgreater{}, and \textless{}122\textgreater{} directions, we find
agreement with the measured transit times across the
$0.60$
mm thick
$(100)$
Si:B substrate only for the \textless{}111\textgreater{} LA and TA modes. The group (and phase) velocities for
the \textless{}111\textgreater{} LA and TA modes are
$v_{L A} =9.400$
km/s and
$v_{T A} =5.110$
km/s, respectively. The computed transit times for the \textless{}111\textgreater{} LA and TA\ modes are
$110$
ns and
$203$
ns, in good agreement with the experimental values,
$110$
ns and
$210$
ns, respectively (Fig. 3). The slight discrepancy with the TA transit time may be due to a
change in the elastic constants due to the Si:B substrate doping. Identical arrival times however,
would be expected for phonons produced along any of the four equivalent \textless{}111\textgreater{} forward
directions. The Bragg scattering
condition for distributed feedback [see Fig. \ref{Figure 1}(c)] is given by:
\begin{equation}2 Q \cos  (\theta ) =N G \label{Bragg condition}
\end{equation}
with
$\theta $
the angle between the phonon wavevector
$\overrightarrow{Q}$
and the DSL
reciprocal lattice vector
$\overrightarrow{G} =\frac{2 \pi }{d} [001]$
, and with
$N$
integer. For the observed \textless{}111\textgreater{} LA and TA\  modes at
$1.04$
THz (frequency verified below), with associated wavelengths
$\Lambda _{L A} =9.35$
nm and
$\Lambda _{T A} =4.91$
nm, respectively, we indeed find       near integer values
$N$
for both LA and TA modes, i.e, 1.00 and 1.90, respectively. Thus, the experimental results
are in excellent agreement with the distributed feedback requirement. A higher threshold
observed for the TA\ mode could be expected since the Bragg
scattering condition is not quite as well satisfied in this case, i.e., 1.90 instead of 2.00.
Interestingly, a [001] LA mode at 1.04 THz would also satisfy the Bragg condition with
$N =1.00$
but this mode is not observed.

As shown in Fig. 3, as the incident FIR estimated power increases, the registered
bolometer signals broaden in time and the amplitudes
saturate. Thus, the bolometer unfortunately was operated in a nonlinear regime in these experiments.
To continue with an analysis, we will assume that the integrated area of a given pulse is
proportional to the absorbed energy. We accordingly assume that the absorbed signal power is
proportional to its integrated pulse area, relative to
that of the integrated weakest laser pulse response (lowest trace Fig. 3). Estimated
absorbed powers were then obtained by multiplying the ratio
of areas by the peak voltage of the weak laser pulse, and dividing by the bolometer responsivity
$\mathcal{R}$
and the post-detection electronics amplification. The estimated incident pump powers are
obtained (numbers shown above traces to the right in Fig. 3) from the absorbed FIR pump powers by
dividing by
$0.15$
(that fraction of the zero-order propagating light reaching the bolometer) and by
$0.016$
, the FIR absorbtance of the bolometer. The latter was calculated using the theory of
electromagnetic wave propagation in
stratified dielectric and metallic media \cite{Born}, using the low-temperature
FIR\ index of refraction of silicon\cite{Loewenstein},
$3.318$, and the resistivity,
$235$
$\mu  \Omega $
-cm, of the biased bolometer. Estimates of the acoustic pulse power within the doping DSL
result from dividing the bolometer's absorbed acoustic power by
$0.30$, the estimated acoustic wave absorptance of the bolometer. The latter is consistent with
dielectric crystal/metal interface studies \cite{Kaplan}.

Figure 4 displays separately, the LA and TA phonon spectra as obtained by Si:B
pieozophonon spectroscopy.
\begin{figure}\centering 
\setlength\fboxrule{0in}\setlength\fboxsep{0in}\fcolorbox[HTML]{FFFFFF}{FFFFFF}{\includegraphics[ width=2.988792in, height=1.9925279999999999in,]{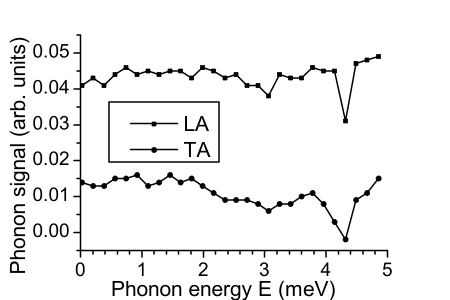}
}
\caption{LA (top) and TA (bottom) phonon signals (points with lines to
guide the eye)\ versus applied [100] uniaxial stress, or phonon
energy (2.73 x 10\textsuperscript {-3} meV/bar). 0.18 meV step size.
}\label{Figure 3}\end{figure}
The averaged ($10^{3}$
single shots) peak of a bolometer response to a phonon pulse, for a time gate of width
$ \Delta t$
positioned at the arrival time for the mode in question, is
plotted versus the Si:B resonant scattering energy
$E$. Here the peak value of each bolometer phonon signal, prior to averaging, was divided by a
the peak value of a separately detected portion of the FIR\ pump
radiation that reflected from a wiremesh beamsplitter and focused onto a pyroelectric detector. As a
result, the averaged bolometer responses shown are in arbitrary units, and the corresponding pump
power was not estimated from the bolometer response in this case. The phonon  energies for both
LA\ and TA signals (the
drop in the signals) are\ identified as
$4.3$
meV, corresponding to
$1.04$
THz, in excellent agreement with the presumed AO\ D4WM
process. On average, the TA signal is weaker
$( \times \frac{1}{4})$
. It is well known from phonon focusing studies in silicon \cite{Wolfe} using incoherent sources that TA phonon signals are stronger than LA signals
along  \textless{}111\textgreater{}. Since our data shown in Figs. 3 and 4 indicates the opposite, we have further
evidence for coherent, and not incoherent, phonon generation.

Figure 5 shows a plot of the estimated LA power versus the
$1.04$
THz estimated pump power.
\begin{figure}\centering 
\setlength\fboxrule{0in}\setlength\fboxsep{0.1in}\fcolorbox[HTML]{FFFFFF}{FFFFFF}{\includegraphics[ width=2.988792in, height=1.9925279999999999in,]{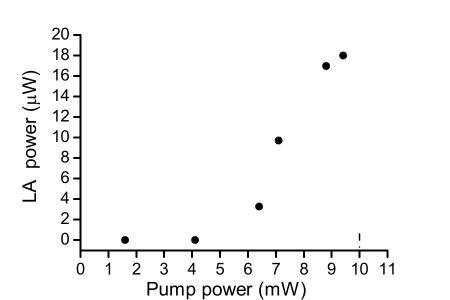}
}
\caption{Estimated LA power versus the estimated pump power of the
FIR\ laser radiation.
}\label{Figure 4}\end{figure}
Here, we have divided the full incident LA\ power by
$4$
to account for power assumed to emanate from four  \textless{}111\textgreater{} equivalent directions. For the
pump power
$P \sim 9.4$
mW and estimated  \textless{}111\textgreater{} LA power
$ \sim 18$
$\mu $
W, we find a conversion efficiency:
$\eta  \sim 2 \times 10^{ -3}$
. The propagating acoustic pulse, contains
$9 \times 10^{ -14}$
J in a volume
$V =v_{L A}  \Delta t A/\sqrt{3} =1.4 \times 10^{ -11}$
cm\textsuperscript {3}. The estimated phonon occupation number in this volume is
$n =1.4 \times 10^{8}$
. The zero point amplitude \cite{Kittel} is given by
$\zeta _{z p} =\sqrt{\hbar  \rho  V^{ \ast } \omega } \sim 3.2 \times 10^{ -18}$
m , where we choose for the initial coherence volume
$V^{ \ast } \sim 1.4 \times 10^{ -15}$
m$^{3}$
, a cylinder formed by the circular projected bolometer area
$A/\sqrt{3}$
, and a coherence length
$\Lambda _{L A}/2$
(due to elastic scattering at
$B$
and
$Sb $
impurities). The avalanche of stimulated LA\ phonons
than corresponds to a classical acoustic wave of amplitude,
$\zeta  =$
$\zeta _{z p} \sqrt{n +1}$
$ \sim 3.7 \times 10^{ -14}$
m. This amplitude corresponds to
$2.7 \times 10^{ -4}$
of the distance between an impurity and the nearest Si atom, or strain,
$S$
$ \sim 3 \times 10^{ -4}$. Picosecond acoustic solitons formed in Si have reported strain levels of this order of
magnitude \cite{Muskens2}. For an active medium
length of
$L =30 \times d \times \sqrt{3} =0.42$
$\mu $m, and assuming a unity zero-point fluctuation, we find a large exponential
gain coefficient:
$g_{L A} =\ln  (n/1)/L$
$ \simeq 4 \times 10^{5}$
cm$^{ -1}$
.

In summary, we report evidence for a self-starting MAPPO operating at
$1.04$
THz.\  A MAPPO may prove useful as a novel source of coherent
acoustic phonons with both compressional and shear waves\  available for
probing material properties at the nanoscale. In principle, a MAPPO \ might also be operated in reverse \cite{Rogovin} as an acoustic wave detector; coherent incident acoustic waves could excite
coherent FIR\ radiation which could be coupled
out using a grating or a prism. Given that stimulated optical 3WM and 4WM
have been used to study quantum correlations between photon
beams\cite{Mei,Lamas-Linares,Grassani}, nonlinear MAPPOs could also provide a
testbed for analogous studies of non-classical acoustic phonon fields.

\begin{acknowledgments}I wish to extend my appreciation for fruitful discussions to K.F. Renk, to E.
Kasper, J. Schulze, and M. Oehme for providing the DSL samples, to A. Lichtenberger for the
NbTiN grating fabrication, and to K. A. Korolev and S. Barber for technical assistance.
\end{acknowledgments}


\begin{thebibliography}{99}
\bibitem {Cahill}D. G. Cahill, P. V. Braun, G. Chen, D. R. Clarke, S. Fan, K. E.
Goodson, P. Keblinski, W. P. King, G. D. Mahan, A. Majumdar, H. J. Maris, S. R., Phillpot, E. Pop,
and L. Shi, Appl. Phys. Rev. \textbf{1}, 011305 (2014).


\bibitem {Tilstra}L.G. Tilstra, A.F.M. Arts, and H.W. de Wijn, Phys. Rev. B \textbf{68}, 144302
(2003).


\bibitem {Ruello}P. Ruello, V.E. Gusev, Ultrasonics \textbf{56}, 21 (2015).


\bibitem {Chen}C-C. Chen, H-M. Huang, T-C. Lu, H-K. Kuo, and C-K.\ Sun, Appl. Phys. Lett. \textbf{100}, 201905 (2012).


\bibitem {Maryam}W. Maryam, A.V. Akimov, R.P. Campion, A.J. Kent, Nature Commun.
\textbf{4}, 2184 (2013).


\bibitem {Kent2}A. J. Kent and R. Beardsley, in \textit{Length-Scale Dependent Phonon
Interactions}, edited by S. L. Shind{\'e} and P. Shrivastava (Springer, NY,
2014), p. 227. Chap. 8.1.


\bibitem {Shinokita}K. Shinokita, K. Reimann, M. Woerner, T. Elsaesser,
R. Hey, and C. Flytzanis, Phys. Rev. Lett. \textbf{116}, 075504 (2016).


\bibitem {Yariv and Pepper}Amnon Yariv and David M. Pepper, Opt. Lett. \textbf{1}, 16 (1977).


\bibitem {Mei}Y. Mei, X. Guo, L. Zhao, and S. Du, Phys. Rev. Lett. \textbf{119}, 150406
(2017).


\bibitem {Brignon}M.J. Damzen, in \textit{Phase Conjugate Laser Optics}, edited by A.
Brignon and J.-P. Huigard (Wiley, Hoboken, 2004), Ch. 11.8, p. 390.


\bibitem {Dohler1}G. H. Dohler, Critical Reviews in Solid State and Materials Sciences,
\textbf{13}(2), 97 (1986).


\bibitem {Ruden} P. Ruden and G.H. Dohler, Solid State Comm. \textbf{45}, 23, 1983.


\bibitem {Gossmann} H.-J. Gossmann and E.F. Shubert, Critical
Reviews in Solid State and Materials Sciences, \textbf{18}(1), 1 (1993).


\bibitem {Tanaka}K. Shinokita, H. Hirori, T. Tanaka, T. Mochizuki, C. Kim, H. Akiyama, L.N.
Pfeiffer, and K.W. West, Phys. Rev. Lett. \textbf{111}, 067401 (2013).


\bibitem {Flytzanis}C. Flytzanis, AIP Conference Proceedings
\textbf{1022}, 471 (2008).


\bibitem {Fossheim}Kristian Fossheim, in \textit{Nonequilibrium Phonon
Dynamics, NATO }\ \textit{ASI}\textit{ }\ \textit{Series}\textbf{124}, edited by Walter E. Bron (Plenum Press, New
York and London, 1984), p. 277.


\bibitem {Xiang}D. Xiang, J. Wu, J. Rottler, and R. Gordon, Nano Lett. \textbf{16}, 3638
(2016).


\bibitem {Rogovin}D. Rogovin, Phys. Rev. A \textbf{41}, 6805 (1990).


\bibitem {Leone1}B. Leone, B. D. Jackson, J. R. Gao, and T. M. Klapwijk, Appl. Phys. Lett.
\textbf{76}(6), 780 (2000).


\bibitem {Li and McCombe}W. J. Li, B.D. McCombe, J. Appl. Phys. \textbf{71}(2), 1038 (1992).


\bibitem {Wilson1}T. E. Wilson, K. A. Korolev, and N. A. Crow, J. Micro/Nanolith.
\textbf{14}, 014501 (2015).


\bibitem {Danilchenko1}B.A. Danilchenko, Cz. Jasiukiewicz, T. Paszkiewicz and S. Wolksi, Acta Phys.
Pol. A \textbf{103}, 325 (2003).


\bibitem {Berberich and Schwarte 1}P. Berberich and M. Schwarte, Z. Phys. B - Condensed Matter \textbf{64}, 1
(1986). We use the static potential constant value of 2.1 eV for Si:B given
in Table 4, and the expression for the ground state splitting energy E as a function of stress X
along [001], given in Table 3. We also make use of the
low-temperature (4 K) elastic constants for Si found in Table 1 of \cite{BerkeMayerWehner}.


\bibitem {Berberich and Kinder}P. Berberich and H. Kinder, J. Phys. Colloques \textbf{42,} C6-374 (1981).


\bibitem {Schwarte and Berberich}M. Schwarte and P. Berberich, J. Phys. C: Solid State Phys \textbf{18}, 3225
(1985).


\bibitem {Froelich1}D. Fr{\"o}hlich and W. Nieswand, Philos. Mag. B \textbf{70}(3), 321 (1994).


\bibitem {Wilson2}T. E. Wilson, Int. J. Infrared and Millimeter Waves \textbf{14}, 303
(1993).


\bibitem {Renk2}K. F. Renk and L. Genzel, Appl. Optics \textbf{1}, 642 (1962).


\bibitem {Gross}C. T. Gross, J. Kiess, A. Mayer, and F. Keilmann, IEEE J. Quantum Electron.
\textbf{QE-23}, 377 (1987).


\bibitem {Woskov1}P. P. Woskov, V. S. Bajaj, M. K. Hornstein, R. J. Temkin, and
R. G. Griffin, IEEE Trans Microw Theory Tech.\textbf{ 53}(6), 1863 (2005).


\bibitem {Auld}B. A. Auld, \textit{Acoustic Waves and Fields, }2nd ed. (Kreiger
Publishing, Malabar, 1990), Vol.1.


\bibitem {BerkeMayerWehner}A. Berke, A.\ P.\ Mayer
and R. K.\ Wehner, J. Phys.
C:\ \textit{Solid State Physics
}\textbf{21}\textit{, }2305 (1988).


\bibitem {Born}M. Born, E. Wolf, \textit{Principles of Optics}, 6th ed. (Pergamon
Press, Oxford, 1980), p. 611.


\bibitem {Loewenstein}Ernest V. Loewenstein, Donald R. Smith, and Robert L. Morgan, Appl. Opt.
\textbf{12}, 398 (1973), p. 405.


\bibitem {Kaplan}S. B. Kaplan, J. Low Temp. Physics \textbf{37}, 343 (1979).


\bibitem {Wolfe}J. P. Wolfe, \textit{Imaging phonons: Acoustic Wave Propagation in
Solids} (Cambridge University Press, Cambridge , 1998). p. 116. The slowness sheet for the
LA mode possesses only convex curvature and cannot produce phonon-focusing caustics.


\bibitem {Kittel}C. Kittel, \textit{Introduction to Solid State Physics}, 7th ed.
(Wiley, New York, 1996), p.108


\bibitem {Muskens2}O.L. Muskens and J.I. Dijkhuis, \textit{Propagation and Diffraction of
Picosecond Acoustic Packets in the Soliton Regime}, in \textit{Optical Solitons},
edited by K. Porsezian and V.C. Kariakose (Springer, Berlin, 2003), p. 391.


\bibitem {Lamas-Linares}A. Lamas-Linares, J.C. Howell and D. Bouwmeester, Nature,
\textbf{412}, 887 (2001).


\bibitem {Grassani}D. Grassani, A. Simbula, S. Pirotta, M. Galli, M. Menotti, N. Harris, T.
Baehr-Jones, M. Hochberg, C. Galland, M. Liscidini, and D. Bajoni, Scientific Reports
\textbf{6}, 23564 (2016).
\end{thebibliography}
\end{document}